\documentclass{pi2005}

\usepackage{amsfonts, amsmath}
\usepackage{cite}

\newtheorem{1}{Proposition}
\newtheorem{2}{Theorem}
\newtheorem{3}{Corollary}
\newtheorem{4}[3]{Corollary}

\slacs{.4ex}

\begin{document}
\title{Design of high-order short-time approximations as
a problem of matching the covariance of a Brownian motion}
\authori{Cristian Predescu}
\addressi{Department of Chemistry and Kenneth S. Pitzer Center for Theoretical Chemistry,\\
University of California, Berkeley, California 94720}
\authorii{}     \addressii{}
%
\headauthor{Cristian Predescu}
\headtitle{A problem of matching the covariance of a Brownian
motion}
\lastevenhead{Cristian Predescu: A problem of matching the
covariance of a Brownian motion}

\maketitle

\begin{abstract}
One of the outstanding problems in the numerical discretization of
the Feynman--Kac formula calls for the design of arbitrary-order
short-time approximations that are constructed in a stable way,
yet only require knowledge of the potential function. In essence,
the problem asks for the development of a functional analogue to
the Gauss quadrature technique for one-dimensional functions. In
PRE \textbf{69} (2004) 056701, it has been argued that the problem
of designing an approximation of order $\nu$ is equivalent to the
problem of constructing discrete-time Gaussian processes that are
supported on finite-dimensional probability spaces and match
certain generalized moments of the Brownian motion. Since Gaussian
processes are uniquely determined by their covariance matrix, it
is tempting to reformulate the moment-matching problem in terms of
the covariance matrix alone. Here, we show how this can be
accomplished.
\end{abstract}

\section{Introduction}

Since its introduction, the Feynman--Kac formula \cite{Fey48,
Kac51, Sim79} has played a fundamental role in the development of
numerical algorithms capable of accounting for the physical
properties of quantum systems made up of distinguishable
particles. When utilized in tandem with the Monte Carlo
integration technique, a powerful method is obtained: the
path-integral Monte Carlo \cite{Cep95}, which is capable of
probing the quantum effects without any untestable approximations.
The Feynman--Kac formula expresses the density matrix of a
thermodynamic system as the expected value of a functional of the
Brownian bridge
\begin{equation}
\label{eq:I1} \rho(x,x';\beta)=\frac{1}{\sqrt{2\pi\sigma^2}}\,
\E^{-(x'-x)^2/2\sigma^2}\mathbb{E}\exp\left\{-\beta\int_{0}^{1}\!
\!  V\Big[x_r(u)+\sigma B_u^0 \Big]\D u\right\}.
\end{equation}
A second formulation of the Feynman--Kac formula is in terms of
the full Brownian motion and reads
\begin{equation}
\label{eq:I1p}
\begin{array}{rcl}
\left<x\left|\E^{-\beta H}\right|\psi\right>&\equiv&\disty
\int_{\mathbb{R}}\rho(x,x';\beta)\psi(x')\D x'=\\[9pt]
&=&\disty \mathbb{E}\left[\exp\Bigl(-\beta\int_{0}^{1}V(x+\sigma
B_u)\D u\Bigr)\psi(x+\sigma B_1)\right],
\end{array}
\end{equation}
where $\psi(x)$ is any square integrable function. In the above,
$\rho(x,x';\beta)$ is the density matrix for a one-dimensional
canonical system characterized by the inverse temperature
$\beta=1/(k_B T)$ and made up of identical particles of mass $m_0$
moving in the potential $V(x)$. The stochastic element that
appears in Eq.~(\ref{eq:I1}), $\{B_u^0,\,0\leq u\leq 1\}$, is a
so-called standard Brownian bridge, defined as follows: if
$\{B_u,\,u\geq 0\}$ is a standard Brownian motion starting at
zero, then the Brownian bridge is the stochastic
process~$\{B_u,\,0 \leq u \leq 1 |\,B_1=0\}$, i.e., a Brownian
motion conditioned on the event~$B_1=0$. As is well known, a
Brownian bridge can be realized as  the process
$\{B_u-uB_1,\,0\leq u\leq 1\}$ \cite{Dur96}. To complete the
description of Eqs.~(\ref{eq:I1}) and~(\ref{eq:I1p}), we have
$x_r(u)=x+(x'-x)u$ and $\sigma=(\hbar^2\beta/m_0)^{1/2}$.

A problem of direct interest to the chemical physicist is the
development of approximations supported on finite-dimensional
probability spaces that have fast convergence for smooth enough
potentials, as measured against the number of evaluations of the
potential function. Most desirably, such approximations should
utilize only the potential function in their construction. For
reasons of stability, they should converge, perhaps at a slower
rate, for all continuous potentials that are bounded from below.
Until recently, the fastest  method available (as order of
convergence) has been the trapezoidal Trotter discrete path
integral method. The technique is usually derived by means of the
Lie--Trotter product formula and an appropriate short-time
high-temperature approximation. The formal asymptotic convergence
of the trapezoidal Trotter method and of  related  techniques was
extensively studied by Suzuki \cite{Suz91, Suz85} for bounded
operators and by Ichinose and Tamura \cite{Ich01, Ich04}, among
others. In particular, results of the last two authors
\cite{Ich04} show that the symmetric Trotter--Suzuki approximation
has optimal convergence $O(1/n^2)$ for sufficiently smooth
potentials as far as pointwise convergence of their integral
kernels is concerned. This type of convergence is also implied in
the present paper.  On the other hand, the non-existence theorem
of Suzuki \cite{Suz91} makes it implausible that faster
convergence can be achieved by utilizing short-time approximations
constructed as functions of the kinetic and potential operators.

Recently, a more general approach has been put forward by the
present author \cite{Pre69-04}, who argued that, for sufficiently
smooth potentials, there might exist direct short-time
high-temperature approximations of arbitrary polynomial
convergence order. Of course, these short-time approximations are
generally not functions of the kinetic and potential operators.
The construction of such approximations is based upon an
``experimental'' theorem on the pointwise convergence of the
integral kernels of the Lie--Trotter product formulas. Although
not rigorously proved, this theorem seems quite plausible. The
short-time approximations considered  are based on carefully
designed finite-dimensional approximations to the Brownian motion
entering the Feynman--Kac formula. Basically, the Brownian motion
is replaced by some discrete-time Gaussian process that is
supported on a finite-dimensional probability space. A set of
functional equations involving some generalized moments of the
Gaussian process have been shown to control the order of
convergence $\nu$. Because the number of equations increases in an
exponential fashion with $\nu$, explicit solutions have been
obtained only for $\nu=3,\,4$.

In the present work, we exploit the fact that both the Brownian
motion and its replacement are Gaussian processes and are,
therefore, uniquely determined by their covariance matrices. We
thus show how to express the functional equations in terms of the
covariance matrices alone. Hopefully, the new equations will prove
easier to utilize in a complete mathematical proof of the
existence of short-time approximations of arbitrary order. In the
appendix, we give a general convergence theorem regarding the
construction of finite-dimensional discrete approximations to the
Feynman--Kac formula.

\section{Statement of the moment-matching problem}

Perhaps one of the oldest techniques for simulating a Brownian
bridge (or Brownian motion) is via random series. As such, let
$\{\lambda_k(\tau)\}_{k \geq 0}$ be any orthonormal basis in
$L^2[0,1]$ such that $\lambda_0(\tau)\equiv1$, let
$$
\Lambda_k(u)=\int_{0}^{u}\lambda_k(\tau)\D\tau \quad \text{for}
\quad k\geq 0
$$
and let $\bar{a}:=\{a_0,a_1,\ldots\}$ be a sequence of independent
identically distributed  standard normal random variables. By the
Ito--Nisio theorem \cite{Kwa92}, the random series
$\sum_{k\geq0}a_k\Lambda_k(u)$ is uniformly convergent almost
surely and equal in distribution with a standard Brownian motion
$B_u$ starting at zero. By the construction of a Brownian bridge
as the process $\{B_u-uB_1,\,0\leq u\leq 1\}$ and the fact that
$\Lambda_0(u)=u$, it follows that
$\sum_{k\geq1}a_k\Lambda_k(u)=\sum_{k\geq0}a_k\Lambda_k(u)-a_0u$
is equal in distribution with a standard Brownian bridge. If
$\Omega$ is the set of all sequences $\bar{a}:=\{a_0,a_1,\ldots\}$
and if
$$
\D P[\bar{a}]=\prod_{k=0}^\infty \D\mu(a_k) \quad \text{with}
\quad \D\mu(z) = (2\pi)^{-1/2}\exp(-z^2/2)\D z
$$
is the probability measure on $\Omega$ associated with the
sequence of independent random variables
$\bar{a}:=\{a_0,a_1,\ldots\}$, then the Feynman--Kac formula given
by Eq.~(\ref{eq:I1}) reads
\begin{equation}
\label{eq:2} \rho(x,x';\beta)=\rho_{fp}(x,x';\beta)\int_{\Omega}\D
P[\bar{a}] \exp\left\{-\beta\int_{0}^{1}V\Bigl[x_r(u)+\sigma
{\textstyle\sum\limits_{k=1}^\infty}a_k \Lambda_k(u)\Bigr]\D
u\right\}.
\end{equation}
Here,
$\rho_{fp}(x,x';\beta)=\exp[-(x'-x)^2/2\sigma^2]/(2\pi\sigma^2)^{1/2}$
is recognized as the density matrix of a free particle. The
alternative formulation given by Eq.~(\ref{eq:I1p}) becomes
\begin{equation}
\label{eq:2p} \rho(x,x';\beta)=\int_{\Omega}\D P[\bar{a}]
\exp\left\{-\beta\int_0^1\!\!V\Bigl[x+\sigma{\textstyle\sum\limits_{k=0}^\infty}
a_k \Lambda_k(u)\Bigr]\D u\right\}\psi(x+\sigma a_0)\,.
\end{equation}
Eqs.~(\ref{eq:2}) and (\ref{eq:2p}) are appropriately called
random series representations of the Feynman--Kac formula in the
chemical-physics literature \cite{Dol84, Pre02}.

Eq.~(\ref{eq:2}) is suggestive of some sort of numerical
approximation to the Feynman--Kac formula, namely
\begin{equation}
\label{eq:3}
\rho_n(x,x';\beta)=\rho_{fp}(x,x';\beta)\int_{\Omega}\D P[\bar{a}]
\exp\left\{-\beta{\textstyle\sum\limits_{i=1}^{n_q}}w_i
V\Bigl[x_r(\theta_i)+\sigma{\textstyle\sum\limits_{k=1}^{n_\nu}}
a_k\Lambda_k(\theta_i)\Bigr]\right\}.
\end{equation}
Here, the non-negative weights $w_i$ (we assume $\sum_i w_i=1$)
and the knots $\theta_i$ define a quadrature technique on the interval
$[0,1]$. It goes without saying that such approximations are
convergent as $n_q\to\infty$ and $n_\nu\to\infty $ under mild
assumptions for the potential $V(x)$: boundness from below and
continuity. In order not to disrupt from the flow of the
presentation, we give the simple proof in the appendix (see
Corollary~\ref{Co:1}). Thus, we are in no shortage of quadrature
formulas. We just want the faster ones.

Another utilization of Eq.~(\ref{eq:3}) is as a short-time
approximation in a Lie--Trotter product of the form
\begin{equation}
\label{eq:4} \rho_n(x,x';\beta)=\int_{\mathbb{R}}\D x_1\ldots
\int_{\mathbb{R}}\D x_n\,
\rho_0^{(\nu)}\left(x,x_1;\frac{\beta}{n+1}\right)\ldots
\rho_0^{(\nu)}\left(x_n,x';\frac{\beta}{n+1}\right).
\end{equation}
The gain in interpreting Eq.~(\ref{eq:3}) as a short-time
approximation is that the requirement that the functions
$\{\Lambda_k(u);\,1\leq k\leq n_\nu\}$ are constructed according
to the Ito--Nisio prescription can be relaxed. We denote this by
utilizing a tilde overscript, so that the short-time approximation
reads
\begin{equation}
\label{eq:5} \begin{array}{rcl}
\rho_0^{(\nu)}(x,x';\beta)&=&\disty
\rho_{fp}(x,x';\beta)\int_{\mathbb{R}}\D\mu(a_1)\ldots
\int_{\mathbb{R}}\D\mu(a_{n_\nu})\times\\[9pt]
&&\disty\times \exp\left\{-\beta\int_0^1V\Bigl[x_r(u)+\sigma
{\textstyle\sum\limits_{k=1}^{n_\nu}}
a_k\tilde{\Lambda}_k(u)\Bigr]\D\omega(u)\right\}.
\end{array}
\end{equation}
The functions $\{\tilde{\Lambda}_k(u);\,1\leq k \leq n_\nu\}$ are
required to be continuous (that is, bounded on the quadrature knots) and vanish at both end points. At this
moment, we should emphasize that this special choice of short-time
approximation is made while bearing in mind its usefulness in
Monte Carlo simulations. Due to some special properties,
especially the availability of the fast sampling algorithm
\cite{Pre71R-05}, the subsequence of Lie--Trotter products with
$n=2^k-1$ is of utmost practical interest. A simple proof of the
convergence of this Lie--Trotter subsequence is given in the
appendix (see Corollary~\ref{Co:2}), again for continuous and
bounded from below potentials.

A second requirement that we ask of the construction given by
Eq.~(\ref{eq:5}) has to do with the symmetry of the density
matrix, which should reflect itself in the symmetry of the
short-time approximation. As such, we require that the discrete
probability measure
\begin{equation}
\label{eq:6}
\D\omega(u)=\sum_{i=1}^{n_q}w_i\delta(u-\theta_i)du
\end{equation}
defining the quadrature technique on $[0,1]$ must be symmetric
about $\frac12$. Also, the finite dimensional process
$\sum_{k=1}^{n_\nu} a_k \tilde{\Lambda}_k(u)$ must be invariant
under the transformation $u'=1-u$. That is, we require the
equality in distribution
\begin{equation}
\label{eq:7}
\sum_{k=1}^{n_\nu}a_k\tilde{\Lambda}_k(u)\stackrel{d}{=}
\sum_{k=1}^{n_\nu} a_k \tilde{\Lambda}_k(1-u)\,.
\end{equation}
The time symmetry of the process
$\sum_{k=1}^{n_\nu}a_k\tilde{\Lambda}_k(u)$ can be enforced, for
example, by restricting the functions
$\{\tilde{\Lambda}_k(u);\,1\leq k \leq n_\nu\}$ to the class of
symmetric and antisymmetric functions. To understand these
requirements, notice that the Hermiticity of the density matrix
stems from the symmetry of the Lebesgue measure on $[0,1]$ as well
as from the time symmetry of the standard Brownian bridge $B_u^0$,
i.e., the fact that $\{B_{1-u}^0,\,0\leq u \leq 1\}$ is a Brownian
bridge equal in distribution to $\{B_{u}^0,\,0\leq u \leq 1\}$.
Because the random sum $\sum_{k=1}^{n_\nu}a_k\tilde{\Lambda}_k(u)$
is intended as a replacement for the Brownian bridge, it is
convenient to introduce the notations
\begin{equation}
\label{eq:8}
\tilde{B}_u^0\equiv\sum_{k=1}^{n_\nu}a_k\tilde{\Lambda}_k(u) \quad
\text{and} \quad \tilde{B}_u \equiv a_0u + \tilde{B}_u^0  \quad
\text{for} \ 0\leq u \leq 1\,.
\end{equation}
Thus, the Brownian bridge (motion) is approximated by a simple
Gaussian process that is supported on a finite-dimensional
probability space. Moreover, only the values of the process for
the discrete times represented by the quadrature knots $\theta_1$,
$\theta_2$, \ldots, $\theta_{n_q}$ are relevant for the construction of the
short-time approximation.

The symbol $(\nu)$ appearing as a superscript in Eq.~(\ref{eq:5})
denotes the order of convergence of the short-time approximation.
This is defined as the largest $\nu$ for which the so-called
convergence operator $T_\nu\psi$ expressed by
\begin{equation}
\label{eq:9}
(T_\nu\psi)(x)=\lim_{\beta\to0^+}\frac{{\textstyle\int_\mathbb{R}}
\left[\rho_0^{(\nu)}(x,x';\beta)-\rho(x,x';\beta)\right]\psi(x')\,
\D x'}{\beta^{\nu+1}}\,.
\end{equation}
is well defined at least on the class of infinitely differentiable
and compactly supported functions $\psi(x)$. In
Ref.~\cite{Pre69-04}, it is claimed but not rigorously
demonstrated that
\begin{equation}
\label{eq:10}
\lim_{n\to\infty}{(n+1)^\nu}\left[\rho_n(x,x';\beta)-\rho(x,x';\beta)\right]=
\beta^{\nu+1}\int_{0}^1\left\langle x\left|\,\E^{-\theta\beta
H}T_\nu \E^{-(1-\theta)\beta H}\right|x'\right\rangle\D\theta\,,
\end{equation}
where $\rho_n(x,x';\beta)$ is defined by Eq.~(\ref{eq:4}). A
rigorous proof of Eq.~(\ref{eq:10}) is beyond the mathematical
abilities of the present author. Based on the more or less formal
arguments presented in the aforementioned reference, it is very
likely that the statement is true. The author would be very
grateful to the mathematically more inclined reader who may want
to investigate the problem and prove or disprove the assertion
(again, the case $n=2^k-1$ suffices for all practical purposes).

Nonetheless, Eq.~(\ref{eq:10}) states that the convergence of the
Lie--Trotter product is as fast as $1/n^\nu$, which explains the
nomenclature regarding the order of convergence. We should
emphasize that merely the existence of the convergence operator
expressed by Eq.~(\ref{eq:9}) sets some constraints on the
smoothness of the potential function. The natural class of
potentials to study the problem of constructing short-time
approximations of arbitrary order is the class of continuously and
infinitely differentiable functions for which
\begin{equation}
\label{eq:11} \frac{1}{\sqrt{2\pi\alpha}}\int_{\mathbb{R}}
\E^{-z^2/(2\alpha)}\left|V^{(k)}(x+z)\right|^j\D z<\infty\,,
\end{equation}
for all $x\in\mathbb{R}$ and $\alpha>0$ and for all integers
$k\geq0$ and $j\geq1$. This condition is necessary in order to
ensure that we recover the original potential $V(x)$, derivatives
$V^{(k)}(x)$, or products of such functions from their Gaussian
transforms, in the limit that $\alpha\to0$ (see Theorem~3 of
Ref.~\cite{Pre02f}). We mention that this class of potentials does
not include some pathological infinitely differentiable and
bounded from below potentials such as $\exp(x^4)$ or even
$\cos[\exp(x^4)]$.

In these conditions, according to Theorem~4 of
Ref.~\cite{Pre69-04}, the convergence operator $T_\nu$ exists if
and only if
\begin{equation}
\label{eq:12}
\mathbb{E}\left[({B}_1)^{j_1}(M_0)^{j_2}({M}_1)^{j_3}\ldots
({M}_{2\mu-2})^{j_{2\mu}}\right]=
\mathbb{E}\left[(\tilde{B}_1)^{j_1}(\tilde{M}_0)^{j_2}
(\tilde{M}_1)^{j_3}\ldots (\tilde{M}_{2\mu-2})^{j_{2\mu}}\right]
\end{equation}
for all $2\mu$-tuples of non-negative integers
$(j_1,j_2,\ldots,j_{2\mu})$ such that $\sum_{k=1}^{2\mu}kj_k=2\mu$
and $1\leq\mu\leq\nu$. The random variables $M_k$ and
$\tilde{M}_k$ are defined by
\begin{equation}
\label{eq:13} M_k \equiv \int_0^1 (B_u)^k\D u\quad\text{and}\quad
\tilde{M}_k \equiv \int_0^1 (\tilde{B}_u)^k\D\omega(u)=
\sum_{i=1}^{n_q}w_i(\tilde{B}_{\theta_i})^k\,,
\end{equation}
respectively.

\section{Reformulation of the generalized moment conditions in terms
of covariance matrices}

Since they are Gaussian processes with continuous paths, the
Brownian motion and its approximation are uniquely determined by
the covariance matrices
\begin{equation}
\label{eq:16} \gamma(u,\tau)=\mathbb{E}(B_u
B_\tau)=u\tau+\sum_{k=1}^\infty \Lambda_k(u)\Lambda_k(\tau)=
\min\{u,\tau\}
\end{equation}
and
\begin{equation}
\label{eq:17}
\tilde{\gamma}(u,\tau)=\mathbb{E}(\tilde{B}_u\tilde{B}_\tau)=
u\tau+\sum_{k=1}^{n_\nu}\tilde{\Lambda}_k(u)\tilde{\Lambda}_k(\tau)\,,
\end{equation}
respectively. As such, at least in principle, the relations given
by Eq.~(\ref{eq:12}) can be formulated in terms of these
covariance matrices alone. In this section, we show how this can
be done.

Let $J_{2\mu}$ denote the set of solutions of the Diophantine
equation $\sum_{k=1}^{2\mu}kj_k=2\mu$. For each $\zeta\in
J_{2\mu}$, we define the integer
\begin{equation}
\label{eq:18}
d(\zeta) = j_3 + j_4 + \cdots + j_{2\mu}
\end{equation}
and
\begin{equation}
\label{eq:19}
n(\zeta)=\, [j_1+j_3+2j_4+\cdots+(2\mu-2)j_{2\mu}]=\mu-j_2-d(\zeta)\,,
\end{equation}
respectively. We also define the differentiation functional
$\mathcal{D}_\zeta$ acting on the space of infinitely
differentiable functions
$f(\lambda_0,\lambda_1,\ldots,\lambda_{d(\zeta)})$ that associates
to each $f$ the following partial derivative evaluated at the
origin
\begin{equation}
\label{eq:20}
\begin{array}{rcl}
\mathcal{D}_{\zeta}f&\equiv&\dfrac{\partial^{j_1}}{\partial\lambda_0^{j_1}}\,
\dfrac{\partial^{j_3}}{\partial\lambda_1\partial\lambda_2\cdots\partial\lambda_{j_3}}\,
\dfrac{\partial^{2j_4}}{\partial\lambda_{j_3+1}^2\partial\lambda_{j_3+2}^2\cdots
\partial\lambda_{j_3+j_4}^2}\,\cdots\\[9pt]
&&\cdots
\dfrac{\partial^{(2\mu-2)j_{2\mu}}}{\partial\lambda_{d(\zeta)-j_{2\mu}+1}^{2\mu-2}
\partial\lambda_{d(\zeta)-j_{2\mu}+2}^{2\mu-2}\cdots
\partial \lambda_{d(\zeta)}^{2\mu-2}}\,f(0, 0, \ldots, 0)\,.
\end{array}
\end{equation}
The differential order of $\mathcal{D}_\zeta$ is
$j_1+j_3+2j_4+\cdots+(2\mu-2)j_{2\mu}=2n(\zeta)$.

The reader will understand the need for this rather cumbersome
notation short\-ly. Going back to Eq.~(\ref{eq:12}), let us notice
that the equality
$$
M_0=\int_0^1 1\D u=1=\int_0^1 1\D\omega(u)=\tilde{M}_0
$$
means that the factors containing $M_0$ and $\tilde{M}_0$ cancel
out. Next, we utilize the identities
$$
(B_1)^{j_1}=j_1!\frac{\D^{j_1}}{\D\lambda^{j_1}}\,\E^{\lambda
B_1}\Bigl|_{\lambda=0} \quad\text{and}\quad M_k=k!\int_0^1\D
u\frac{\D^{k}}{\D\lambda^{k}}\,\E^{\lambda B_u}\Bigl|_{\lambda=0}
$$
to express the first factor in Eq.~(\ref{eq:12}) as
\begin{equation}
\label{eq:21}
\begin{array}{l}
\disty j_1!(1!)^{j_3}\cdots \bigl((2\mu-2)!\bigr)^{j_{2\mu}}
\mathbb{E}\biggl[\left(\frac{D^{j_1}}{\D\lambda^{j_1}}\,
\E^{\lambda B_1}\Bigl|_{\lambda=0}\right) \left(\int_0^1\D
u\,\frac{\D}{\D\lambda}\,\E^{\lambda
B_u}\Bigl|_{\lambda=0}\right)^{j_3}\cdots\\[12pt]
\disty\hskip40mm
\cdots \left(\int_0^1\D
u\,\frac{\D^{2\mu-2}}{\D\lambda^{2\mu-2}}\,\E^{\lambda
B_u}\Bigl|_{\lambda=0}\right)^{j_{2\mu}}\bigg].
\end{array}
\end{equation}
Expanding the parenthesis and interchanging the order of
integration and differentiation, we obtain the result
$$
j_1!(1!)^{j_3}\cdots\bigl((2\mu-2)!\bigr)^{j_{2\mu}}D_{\zeta}f_{\zeta}\,,
$$
where
$$
f_\zeta(\lambda_0,\lambda_1,\ldots,\lambda_{d(\zeta)})= \int_0^1\D
u_1\cdots \int_0^1\D u_{d(\zeta)}
\mathbb{E}\exp\biggl(\,{\textstyle\sum\limits_{i=0}^{d(\zeta)}}\lambda_i
B_{u_i}\biggr)
$$
and $u_0=1$.

To evaluate the function $f_\zeta$, one may utilize any random
series and compute
$$
\begin{array}{l}
\disty
\mathbb{E}\exp\biggl(\,{\textstyle\sum\limits_{j=0}^{d(\zeta)}}\lambda_i
B_{u_i}\biggr)=\mathbb{E}\exp\biggl(\,{\textstyle\sum\limits_{i=0}^{d(\zeta)}
\lambda_i\sum\limits_{k=0}^\infty} a_k\Lambda_k(u_i)\biggr)=\\[12pt]
\disty\qquad= \prod_{k=0}^\infty
\left\{\int_{\mathbb{R}}\D\mu(z)\exp\biggl(z{\textstyle\sum\limits_{i=0}^{d(\zeta)}}
\lambda_i\Lambda_k(u_i)\biggr)\right\}=
\prod_{k=0}^{\infty}\exp\left\{\frac12\,\biggl[\,{\textstyle\sum\limits_{i=0}^{d(\zeta)}}
\lambda_i\Lambda_k(u_i)\biggr]^2 \right\}.
\end{array}
$$
The last term equals
$$
\exp\left\{\frac{1}{2}{\textstyle\sum\limits_{i=0}^{d(\zeta)}
\sum\limits_{j=0}^{d(\zeta)}\lambda_i\lambda_j
\biggl(\,\sum\limits_{k=0}^\infty}
\Lambda_k(u_i)\Lambda_k(u_j)\biggr)\right\}=
\exp\left\{\frac{1}{2}{\textstyle\sum\limits_{i=0}^{d(\zeta)}
\sum\limits_{j=0}^{d(\zeta)}}\lambda_i\lambda_j\gamma(u_i,u_j)\right\}.
$$
Therefore, the explicit expression of the function $f_\zeta$ is
$$
f_\zeta(\lambda_0,\lambda_1,\ldots,\lambda_{d(\zeta)})= \int_0^1\D
u_1\cdots\int_0^1\D u_{d(\zeta)}
\exp\biggl\{\,\frac{1}{2}{\textstyle\sum\limits_{i=0}^{d(\zeta)}
\sum\limits_{j=0}^{d(\zeta)}}\lambda_i\lambda_j\gamma(u_i,u_j)\biggr\}.
$$
We remind the reader that $u_0=1$.

The function $f_\zeta$ can be replaced by some
$(1+d(\zeta))$-dimensional polynomial of degree $2n(\zeta)$.
Indeed, starting from the expansion
$$
\exp\biggl\{\frac{1}{2}{\textstyle\sum\limits_{i=0}^{d(\zeta)}
\sum\limits_{j=0}^{d(\zeta)}}\lambda_i\lambda_j\gamma(u_i,u_j)\biggr\}=
\sum_{k=0}^\infty\frac{1}{2^kk!}\biggl[\,{\textstyle\sum\limits_{i=0}^{d(\zeta)}
\sum\limits_{j=0}^{d(\zeta)}}\lambda_i\lambda_j\gamma(u_i,u_j)\biggr]^k
$$
one observes that, upon the action of $D_\zeta$, the differential
order of which is $2n(\zeta)$, only the term with $k=n(\zeta)$
survives. The terms of lower degree are washed out by the process
of differentiation, whereas the terms of higher degrees cancel
when the respective derivatives are evaluated at the origin. Thus,
the left-hand side of Eq.~(\ref{eq:12}) is given by the expression
\begin{equation}
\label{eq:22}
\frac{j_1!(1!)^{j_3}\cdots\bigl((2\mu-2)!\bigr)^{j_{2\mu}}}{2^{n(\zeta)}n(\zeta)!}\,D_{\zeta}f_{\zeta}\,,
\end{equation}
where
\begin{equation}
\label{eq:23}
f_\zeta(\lambda_0,\lambda_1,\ldots,\lambda_{d(\zeta)})=
\int_{0}^1\D u_1\cdots\int_0^1\D u_{d(\zeta)}
\biggl[\,{\textstyle\sum\limits_{i=0}^{d(\zeta)}
\sum\limits_{j=0}^{d(\zeta)}}\lambda_i\lambda_j\gamma(u_i,u_j)\biggr]^{n(\zeta)}.
\end{equation}

In an analogue manner, one demonstrates that the right-hand side
of Eq.~(\ref{eq:12}) can be written as
\begin{equation}
\label{eq:24}
\frac{j_1!(1!)^{j_3}\cdots\bigl((2\mu-2)!\bigr)^{j_{2\mu}}}{2^{n(\zeta)}n(\zeta)!}\,D_{\zeta}\tilde{f}_{\zeta}\,,
\end{equation}
where
\begin{equation}
\label{eq:25}
\tilde{f}_\zeta(\lambda_0,\lambda_1,\ldots,\lambda_{d(\zeta)})=
\int_{0}^1\D\omega(u_1)\cdots\int_0^1\D\omega(u_{d(\zeta)})
\biggl[\,{\textstyle\sum\limits_{i=0}^{d(\zeta)}
\sum\limits_{j=0}^{d(\zeta)}}\lambda_i\lambda_j\tilde{\gamma}(u_i,u_j)\biggr]^{n(\zeta)}.
\end{equation}
Comparing Eqs.~(\ref{eq:22}) and (\ref{eq:24}), we see that
Eq.~(\ref{eq:12}) is equivalent to the equality
\begin{equation}
\label{eq:26}
D_{\zeta}{f}_{\zeta}=D_{\zeta}\tilde{f}_{\zeta}
\end{equation}
with $f_\zeta$ and $\tilde{f}_\zeta$ defined by Eqs.~(\ref{eq:23})
and (\ref{eq:25}), respectively.

As Eqs.~(\ref{eq:22}) and (\ref{eq:24}) show, the polynomials
$f_\zeta$ and $\tilde{f}_\zeta$ depend only on some integers
$d(\zeta)$ and $n(\zeta)$ that have the property
$$
d(\zeta)+n(\zeta)\leq\mu\leq\nu\,.
$$
Thus, a sufficient condition that the equality expressed by
Eq.~(\ref{eq:26}) holds for all $\zeta\in J_{2\mu}$ and
$1\leq\mu\leq\nu$ is that the equality of polynomials
\begin{equation}
\label{eq:27}
\begin{array}{l}
\disty \tilde{f}_{n,d}(\lambda_0,\lambda_1,\ldots,\lambda_d)\equiv
\int_{0}^1\D\omega(u_1)\cdots\int_0^1\D\omega(u_d)
\biggl[\,{\textstyle\sum\limits_{i=0}^d
\sum\limits_{j=0}^d}\lambda_i\lambda_j\tilde{\gamma}(u_i,u_j)\biggr]^n=\\[9pt]
\disty\qquad\qquad= \int_0^1\D u_1\cdots\int_0^1\D u_{d}
\biggl[\,{\textstyle\sum\limits_{i=0}^d
\sum\limits_{j=0}^d}\lambda_i\lambda_j\gamma(u_i,u_j)\biggr]^n\equiv
f_{n,d}(\lambda_0,\lambda_1,\ldots,\lambda_d)
\end{array}
\end{equation}
holds for all integer $d$ and $n$ such that $d+n\leq\nu$. If
$d_1\leq d_2$ then the polynomial of dimension $1+d_1$ is a
particular case of the polynomial of dimension $1+d_2$, with the
last $d_2-d_1$ lambda's set to zero. Thus, it is enough to check
the above equality for the cases with $d+n=\nu$.

Let us show that checking the equality expressed by
Eq.~(\ref{eq:27}) for all integers $n$ and $d$ such that $n+d=\nu$
is also a necessary condition. Notice that the polynomials
appearing in Eq.~(\ref{eq:27}) are symmetric under the permutation
of the variables $\lambda_1$, $\lambda_2$, \ldots, $\lambda_d$.
This is due to the symmetry of the covariance matrices
$\gamma(u,\tau)$ and $\tilde{\gamma}(u,\tau)$. In these
conditions, the differentiation functional given by Eq.~(\ref{eq:20})
is the most general differential expression of order $2n$,
provided that  we let the index
$\zeta=\{j_1,j_2,j_3,j_4,\ldots,j_{2\nu}\}$ lie in the set $S$
that contains all possible indexes for which $j_2=0$,
$j_3+j_4+\cdots+j_{2\nu}=d$, and
$j_1+j_3+2j_4+\cdots+(2\nu-2)j_{2\nu}=2n$. Following some
permutation of the variables $\lambda_1$, $\lambda_2$, \ldots,
$\lambda_d$, any other differential expression can be obtain from
and, by the aforementioned symmetry of polynomials, is equal to a
differential expression of the type given by Eq.~(\ref{eq:20}) for
some $\gamma\in S$. Therefore, the equality $\mathcal{D}_\zeta
f_{n,d}=\mathcal{D}_\zeta\tilde{f}_{n,d}$ for all $\zeta\in S$
implies $f_{n,d}=\tilde{f}_{n,d}$. It remains to prove that, as
$\zeta$ spans $J_{2\nu}$, it also spans $S$. Clearly, any index
$\zeta\in S$ has the property $j_1+2j_2+3j_3+\cdots+2\nu j_{2\nu}=
2(n+d)=2\nu$. Therefore, $S\subset J_{2\nu}$ and the claim of
necessity is proved.

We summarize the results of the paper in the following
proposition.
\begin{1}
\label{Pr:1} A short-time approximation of the type given by
Eq.~(\ref{eq:5}) has convergence order $\nu$ if and only if the
equality  of polynomials expressed by Eq.~(\ref{eq:27}) holds for
all integers $d$ and $n$ such that $d+n=\nu$.
\end{1}

The above proposition can be utilized together with the
multinomial formula to generate the necessary conditions that the
trial covariance matrix must satisfy in order for a short-time
approximation to have convergence order $\nu$. Most likely, more
useful statements can be obtained by the mathematically more
inclined reader. Nevertheless, we have achieved our goal of
formulating the conditions in terms of the covariance matrices
alone.

\bigskip
\noindent {\small This work was supported in part by the National
Science Foundation Grant No. CHE-0345280 and by the Director,
Office of Science, Office of Basic Energy Sciences, Chemical
Sciences, Geosciences, and Biosciences Division, U.S. Department
of Energy under Contract No. DE-AC02-05CH11231.}

\section*{Appendix}

In this appendix, we establish an easily provable theorem that is
useful in many situations and provides the basic mathematical
background for the discretization of the Feynman--Kac formula. We
then utilize the theorem to demonstrate several statements made in
the text.

To begin with, let us formalize what we mean by quadrature rules.
By a quadrature scheme on $[0,1]$, we understand a sequence of
pairs of vectors of variable lengths $q_n$, namely the weights
$\{w_{n,1},w_{n,2},\ldots,w_{n,q_n}\}$ and the knots $0\leq
u_{n,1}<u_{n,2}<\cdots<u_{n,q_n}\leq1$, constructed such that the
$w_{n,i}$'s are non-negative  for all $1\leq i\leq q_n$ and
$n\geq1$ and such that
\begin{equation}
\label{eq:A1} \lim_{n\to\infty}\sum_{i=1}^{q_n}w_{n,i}h(u_{n,i})=
\int_0^1h(u)\D u
\end{equation}
for all continuous functions $h:[0,1]\to\mathbb{R}$. The
mid-point, trapezoidal, Simpson, and Gauss--Legendre rules are
well-known examples of quadrature schemes. The non-negativity of
the weights is a stability requirement that simply says that it is
not all right to get a negative answer if a positive
function is integrated.

Let $\{S_{n}(\omega;u);\,0\leq u\leq1\}_{n\geq1}$ be a family of
random processes with continuous paths supported on a probability
space $(\Omega,\mathcal{F},P)$ and such that $S_{n}(\omega;u)$
converges to some limit $B_u^0(\omega)$ uniformly in $u$,
$P$-almost surely. Assume that the limit $B_u^0(\omega)$ is equal
in distribution to a Brownian bridge. Given an arbitrary
quadrature rule, we can define a sequence of approximations to the
density matrix by the prescription
\begin{equation}
\label{eq:A2} \rho_n(x,x';\beta)=\rho_{fp}(x,x';\beta)\mathbb{E}
\exp\biggl\{-\beta{\textstyle\sum\limits_{i=1}^{q_n}}w_{n,i}V
\left[x_r(u_{n,i})+\sigma S_{n}(\omega;u_{n,i})\right]\biggr\},
\end{equation}
for $n \geq 1$. We shall call the above prescription the
\textit{standard discretization of the Feynman--Kac formula}. The
nomenclature is motivated by the following fundamental theorem.

\begin{2}
\label{Th:1} If $V(x)$ is continuous and bounded from below, then
the sequence $\rho_n(x,x';\beta)$ is bounded by
$C(\beta)\rho_{fp}(x,x';\beta)$, for some positive
$C(\beta)<\infty$, and
\begin{equation}
\rho_n(x,x';\beta)\to\rho(x,x';\beta)\,,\ \text{as}\ n\to\infty\,.
\end{equation}
\end{2}

\noindent \textit{Proof}. Let $V_0$ be a non-positive lower bound
for $V(x)$. Eq.~(\ref{eq:A1}) specialized for $h(x)=1$ says that
the sequence $\sum_{i=1}^{q_n}w_{n,i}$ is convergent, thus
bounded, say by $c>0$. Then the integrand of Eq.~(\ref{eq:A2}) is
bounded from above by  $C(\beta)=\exp(-c\beta V_0)$ and we have
\begin{equation}
\label{eq:A3p} \rho_n(x,x';\beta)\leq
C(\beta)\rho_{fp}(x,x';\beta)\,.
\end{equation}
The remainder of the theorem follows from the dominated
convergence theorem and the Feynman--Kac formula, as soon as we
prove that
\begin{equation}
\label{eq:A3}
\lim_{n\to\infty}\sum_{i=1}^{q_n}w_{n,i}V\left[x_r(u_{n,i})+
\sigma S_n(\omega;u_{n,i})\right]=\int_0^1V[x_r(u)+\sigma
B_u^0(\omega)]\D u\,, \ a.s.
\end{equation}

As such, let $\omega\in\Omega$ be given. Since $x_r(u)+\sigma
B_u^0(\omega)$ is almost surely continuous in $u$ --- as uniform
limit of continuous functions $x_r(u)+\sigma S_n(\omega;u)$ --- it
is bounded by some constant $M>0$. In fact, by uniform
convergence, we have $\left|x_r(u)+\sigma S_n(\omega;u)\right|\leq
2M$  for large enough $n$. Since $V(x)$ is continuous, it is
uniformly continuous on the compact set $|x|\leq2M$. Consequently,
given $\epsilon>0$, there is $\eta>0$ such that
$|V(x)-V(y)|<\epsilon$ whenever $|x-y|<\eta$. Nevertheless, by
uniform convergence,
$$
\left|\left[x_r(u)+\sigma S_n(\omega;u)\right]-\left[x_r(u)+
\sigma B_u^0(\omega)\right]\right|<\eta
$$
for large enough $n$ and so,
$$
\left|V\left[x_r(u)+\sigma S_n(\omega;u)\right]-V\left[x_r(u)+
\sigma B_u^0(\omega)\right]\right|<\epsilon\,.
$$
 Since $\sum_{i=1}^{q_n}w_{n,i}<c$, it follows that
\begin{equation}
\label{eq:A4} \left|\sum_{i=1}^{q_n}w_{n,i}V\left[x_r(u_{n,i})+
\sigma S_n(\omega;u_{n,i})\right]-
\sum_{i=1}^{q_n}w_{n,i}V\left[x_r(u_{n,i})+\sigma
B_{u_{n,i}}^0(\omega)\right]\right|<c\epsilon\,.
\end{equation}
By Eq.~(\ref{eq:A1}) and the continuity of $V\left[x_r(u)+\sigma
B_u^0(\omega)\right]$ as a function in $u$, we also have
\begin{equation}
\label{eq:A5} \left|\int_0^1V[x_r(u)+\sigma B_u^0(\omega)]\D u-
\sum_{i=1}^{q_n}w_{n,i}V\left[x_r(u_{n,i})+\sigma
B_{u_{n,i}}^0(\omega)\right]\right|<\epsilon
\end{equation}
for $n$ sufficiently large. Combining Eqs.~(\ref{eq:A4})
and~(\ref{eq:A5}), we obtain
$$
\left|\sum_{i=1}^{q_n}w_{n,i}V\left[x_r(u_{n,i})+ \sigma
S_n(\omega;u_{n,i})\right]-\int_0^1V[x_r(u)+\sigma
B_u^0(\omega)]\D u\right|<(1+c)\epsilon\,.
$$
Since $\epsilon$ is arbitrary, the  almost sure convergence
appearing in Eq.~(\ref{eq:A3}) is demonstrated and the proof of
the theorem is concluded. \hspace{\stretch{1}} $\Box$

\noindent
\textit{Observation.} Because $\rho_n(x,x';\beta)$ is
bounded by $C(\beta)\rho_{fp}(x,x';\beta)$, the dominated
convergence theorem and the above theorem also imply convergence
in the strong topology, that is,
$$
\int_{\mathbb{R}}\rho_n(x,x';\beta)\psi(x')\D x'\to
\int_{\mathbb{R}}\rho(x,x';\beta)\psi(x')\D x'
$$
for all square integrable $\psi(x)$. In fact, by choosing the
sequence of processes $S_n(\omega,n)$ to be constant and equal in
distribution to $B_u^0$, Th.~\ref{Th:1} produces various versions
of the Trotter convergence theorem for various quadrature rules.

As everywhere else in this paper, in the following, it is
understood that the potential $V(x)$ is continuous and bounded
from below.
\begin{3}
\label{Co:1} If $\rho_n(x,x';\beta)$ is defined by
Eq.~(\ref{eq:3}), then $\rho_n(x,x';\beta)\to\rho(x,x';\beta)$, as
$n\to\infty$.
\end{3}
\textit{Proof.} Follows from Th.~\ref{Th:1} and the uniform
convergence of the random series $\sum_{k\geq1}a_k\Lambda_k(u)$ to
a Brownian bridge, as guaranteed by the Ito--Nisio theorem.
\hspace{\stretch{1}} $\Box$

\begin{4}
\label{Co:2} If $\rho_n(x,x';\beta)$ is defined by
Eq.~(\ref{eq:4}), then
$\rho_{2^k-1}(x,x';\beta)\to\rho(x,x';\beta)$, as $k\to\infty$.
\end{4}
\textit{Observation.} The proof we construct for the subsequence
$n=2^k-1$ is based on a special form of the Lie--Trotter product
that is called the L\'evy--Ciesielski form. The information from
the paragraphs below is taken from Ref.~\cite{Pre69-04}. In fact,
for $n=2^k-1$, any Lie--Trotter product can be put in a
L\'evy--Ciesielski form, as shown in Ref.~\cite{Pre71R-05}. This
form has important advantages when it comes to the Monte Carlo
implementation, such as fast computation of paths and fast
sampling. For this reason, we can restrict our attention to the
subsequence $n=2^k-1$, without any loss of generality for actual
applications.

\noindent \textit{Proof of the corollary.} Let $\{a_{l,j};\,1\leq
l\leq k,\,1\leq j\leq 2^{l-1}\}$ and $\{b_{l,j};\,1\leq l\leq q,\,
1\leq j\leq 2^{k}\}$ be two independent sets of i.i.d. standard
normal variables. Let $\{F_{l,j}(u);\,l\geq1,1\leq j\leq2^{l-1}\}$
be the system of Schauder functions on the interval $[0,1]$. The
Schauder functions are the primitives of the Haar $L^2([0,1])$
wavelet bases and can be generated by contractions and
translations as follows. Let $F_{1,1}(u):\mathbb{R}\to\mathbb{R}$
be defined by
\begin{equation}
\label{eq:1.89} F_{1,1}(u)=\left\{\begin{array}{rl}
u\,,\quad&u\in[0,1/2]\,,\\ 1-u\,,\quad&u\in(1/2,1]\,,\\
0\,,\quad&\text{elsewhere}\,.
\end{array}\right.
\end{equation}
Then,
\begin{equation}
\label{eq:1.91} F_{l,j}(u)= 2^{-(l-1)/2} F_{1,1}(2^{l-1}u-j+1)
\end{equation}
for all $l\geq1$ and $1\leq j\leq2^{l-1}$. Extend the functions
$\{\tilde{\Lambda}_l(u);\,1\leq l\leq n_\nu\}$ outside the
interval $[0,1]$ by setting them to zero [the same way the first
Schauder function $F_{1,1}(u)$ was extended to the whole real axis
in Eq.~(\ref{eq:1.89})] and define
\begin{equation}
\label{eq:1.92} G_{l,j}(u)= 2^{-k/2}\tilde{\Lambda}_l(2^k u-j+1)
\end{equation}
for $1\leq l\leq n_\nu$ and $1\leq j\leq2^k$.

Let $\D\omega_k(u)$ denote the discrete measure associated with
the quadrature scheme  specified by the $n_q2^k$ (not necessarily
different) quadrature knots
\begin{equation}
\label{eq:II18} u'_{i,j}=2^{-k}(\theta_i+j-1)\,,\quad 1\leq i\leq
n_q\,,\;\; 1\leq j\leq 2^k
\end{equation}
and the corresponding weights
\begin{equation}
\label{eq:II19} w'_{i,j}=2^{-k}w_i\,.
\end{equation}
The new quadrature knots $u'_{i,j}$ are obtained by contractions
and translations of the original knots $\theta_i$.

With the convention that $a_{l,2^{l-1}+1}=0$ and $b_{l,2^k+1}=0$
for all $l\in\overline{1,k}$, we have
\begin{eqnarray}
\label{eq:II17} \nonumber
\frac{\rho_n(x,x';\beta)}{\rho_{fp}(x,x';\beta)}&=&
\int_{\mathbb{R}}\D a_{1,1}\ldots\int_{\mathbb{R}}\D a_{k,2^{k-1}}
\left(2\pi\right)^{-n/2}
\exp\biggl(-\frac12{\textstyle\sum\limits_{l=1}^k
\sum\limits_{j=1}^{2^{l-1}}a_{l,j}^2}\biggr)\times\\
&&\times\nonumber \int_{\mathbb{R}}\D b_{1,1}\ldots
\int_{\mathbb{R}}\D b_{n_\nu,2^k}\left(2\pi\right)^{-(n+1)n_\nu/2}
\exp\biggl(-\frac12{\textstyle\sum\limits_{l=1}^{n_\nu}
\sum\limits_{j=1}^{2^{k}}} b_{l,j}^2\biggr)\times\\
&&\times \exp\biggl\{-\beta\int_0^1V\biggl[x_r(u)+\sigma
{\textstyle\sum\limits_{l=1}^k}
a_{l,[2^{l-1} u]+1}\,F_{l,[2^{l-1} u]+1}(u)+\nonumber\\
&&\qquad\qquad+\sigma{\textstyle\sum\limits_{l=1}^{n_\nu}}
b_{l,[2^ku]+1}\,G_{l,[2^k u]+1}(u)\biggr]\D\omega_k(u)\biggr\},
\end{eqnarray}
where $[2^{l-1}u]$ and $[2^ku]$ are the integer parts of
$2^{l-1}u$ and $2^k u$, respectively.

Let us verify that the sequence of discrete measures
$d\omega_k(u)$ defines a quadrature scheme. If
$h:[0,1]\to\mathbb{R}$ is a continuous functions then, by its
uniform continuity, for any $\epsilon>0$, there is $\eta>0$ such
that $|h(u)-h(\tau)|<\epsilon$ whenever $|u-\tau|<\eta$. Now, pick
$k$ large enough that $1/2^k<\eta$. The Lebesgue integral of $h$
over $[0,1]$ can be broken in $2^k$ smaller parts over the
intervals $[(j-1)/2^k,j/2^k]$, for $1\leq j\leq2^k$. Since the
length of these intervals is smaller than $\eta$, we have both
\begin{equation}
\label{eq:41} \left|\int_{(j-1)/2^{k}}^{j/2^{k}}h(u)\D u-
2^{-k}h(j/2^k)\right|<2^{-k}\epsilon
\end{equation}
and
\begin{equation}
\label{eq:42}
\left|\sum_{i=1}^{n_q}w'_{i,j}h\left(u'_{i,j}\right)-
2^{-k}h(j/2^k)\right|<2^{-k}\epsilon\,.
\end{equation}
The latter inequality is true because the points $u'_{i,j}$ are in
the interval $[(j-1)/2^k,j/2^k]$ and because
$$
\sum_{i=1}^{n_q}w'_{i,j}=2^{-k}\sum_{i=1}^{n_q}w_i=2^{-k}\,.
$$
From Eqs.~(\ref{eq:41}) and~(\ref{eq:42}), we learn that
\begin{equation}
\label{eq:43} \left|\int_{(j-1)/2^{k}}^{j/2^{k}}h(u)\D u-
\sum_{i=1}^{n_q}w'_{i,j}h\left(u'_{i,j}\right)\right|<2\cdot2^{-k}\epsilon
\end{equation}
and, by summing over all $j$, we obtain
\begin{equation}
\label{eq:44}
\sum_{j=1}^{2^k}\left|\int_{(j-1)/2^{k}}^{j/2^{k}}h(u)\D u-
\sum_{i=1}^{n_q}w'_{i,j}h\left(u'_{i,j}\right)\right|<2\epsilon\,.
\end{equation}
The left-hand side of Eq.~(\ref{eq:44}) is clearly larger than
$$
\left|\sum_{j=1}^{2^k}\left[\int_{(j-1)/2^{k}}^{j/2^{k}}h(u)\D u-
\sum_{i=1}^{n_q}w'_{i,j}h\left(u'_{i,j}\right)\right]\right|=
\left|\int_0^1 h(u)\D u-\int_0^1 h(u)\D\omega_k(u)\right|.
$$
Since $\epsilon$ is arbitrary, it follows that
$$
\lim_{k\to\infty}\int_0^1 h(u)\D\omega_k(u)=\int_0^1 h(u)\D u
$$
and since $h$ is arbitrary, it follows that the sequence of
discrete measures $d\omega_k(u)$  defines a quadrature scheme.

The first part of the random series appearing in
Eq.~(\ref{eq:II17}) is the L\'evy--Ciesielski series, which
converges uniformly to a Brownian bridge. The corollary readily
follows from the result in the preceding paragraph and
Th.~\ref{Th:1}, if we prove that the tail series
\begin{equation}
\label{eq:45}
T_k(\bar{b};u)=\sum_{l=1}^{n_\nu}b_{l,[2^ku]+1}\,G_{l,[2^ku]+1}(u)
\end{equation}
converges to zero uniformly almost surely, as $k\to\infty$ (in
other words, if our correction term does not ruin the uniform
convergence of the L\'evy--Ciesielski series). Let $M>0$ be a
common bound for the functions $\{\tilde{\Lambda}_k(u);\,1\leq k
\leq n_\nu\}$. Then, with the help of Eq.~(\ref{eq:1.92}), we compute
$$
\biggl(\,\max_{0\leq u\leq1}\left|T_k(\bar{b};u)\right|\biggr)^4
\leq \frac{{M}^4}{2^{2k}} \max_{l,j}|b_{l,j}|^4 \leq
\frac{{M}^4}{2^{2k}} \sum_{l = 1}^{n_\nu} \sum_{j = 1}^{2^k}
b_{l,j}^4\,.
$$
Taking the expected value, we get
$$
\mathbb{E}\biggl(\max_{0\leq u \leq
1}\left|T_k(\bar{b};u)\right|\biggr)^4\leq
\frac{{M}^4}{2^{2k}}(3n_\nu 2^k)=\frac{3n_\nu{M}^4}{2^k}\,.
$$

Now, Chebyshev's inequality produces
$$
P\biggl(\,\max_{0\leq u\leq1}\left|T_k(\bar{b};u)\right|\geq
\epsilon\biggr)\leq
\frac{1}{\epsilon^4}\mathbb{E}\biggl(\,\max_{0\leq u \leq1}
\left|T_k(\bar{b};u)\right|\biggr)^4 \leq
\frac{3n_\nu{M}^4}{\epsilon^42^{k}}
$$
and so,
$$
\sum_{k=1}^\infty P\biggl(\,\max_{0\leq u \leq1}
\left|T_k(\bar{b};u)\right|\geq\epsilon\biggr)\leq
\frac{3n_\nu{M}^4}{\epsilon^4}\sum_{k=1}^\infty\frac{1}{2^k}=
\frac{3n_\nu{M}^4}{\epsilon^4}<\infty\,.
$$
The first Borel--Cantelli lemma implies that
$$
P\biggl(\,\max_{0\leq u\leq1}\left|T_k(\bar{b};u)\right|\geq
\epsilon\;\;\mathrm{i.o.}\biggr)=0\,,
$$
which means that, with probability one, there is a rank $K\geq1$
such that
$$
\max_{0\leq u\leq1}\left|T_k(\bar{b};u)\right|<\epsilon\,,\quad
\forall\,k\geq K\,.
$$
Letting $\epsilon$ go to zero through the countable sequence
$\epsilon_j=1/j$, we obtain the almost sure uniform convergence to
zero of the tail series $T_k(\bar{b};u)$. The proof of the
corollary is concluded. \hspace{\stretch{1}} $\Box$

\end {document}